\documentstyle[prl,aps,twocolumn,epsfig]{revtex}
\newcommand{\be}{\begin{equation}}
\newcommand{\ee}{\end{equation}}
\newcommand{\bdis}{\begin{displaymath}}
\newcommand{\edis}{\end{displaymath}}

\begin{document}
\bibliographystyle{prsty}  

\title{Universal statistics of  non-linear energy transfer
in turbulent models}
\author{ R. Benzi$^{1}$, L. Biferale$^{2}$,  and E.Trovatore$^{3}$ }
\address{$^1$ AIPA, Via Po 14, 00100 Roma, Italy\\
$^2$  Dept. of Physics, University of Tor Vergata,
Via della Ricerca Scientifica 1, I-00133 Roma, Italy\\ 
$^{3}$  INFM-Dept. of Physics, University of Cagliari,
Via Ospedale 72, I-09124, Cagliari, Italy}
\date{\today}
\maketitle
\begin{abstract}
A class of shell models for turbulent energy transfer 
at varying the inter-shell separation, $\lambda$, is investigated.
 Intermittent 
corrections in the  continuous limit of infinitely 
close shells ($\lambda \rightarrow 1$) have been measured.
Although the model becomes, in this limit, non-intermittent,
 we found 
universal aspects of the velocity statistics which can be interpreted
in the framework of log-poisson distributions, as proposed 
by She and Waymire (1995, {\it Phys. Rev. Lett.} {\bf 74}, 262). 
We suggest that non-universal aspects of intermittency can be
adsorbed in the parameters describing  statistics and  properties
of the most singular structure. On the other hand, 
universal aspects can be found
by looking at corrections to the  monofractal scaling of the 
most singular structure. Connections with similar results reported
in other shell models investigations and in real turbulent flows
are discussed.
\end{abstract}
\pacs{PACS numbers: 47.27.Eq}
  
One of the basic questions in understanding the physics of
fully developed three dimensional turbulence is the  
dynamical mechanism characterizing 
the  energy transfer from large to small scales.
According to the Kolmogorov theory (K41) of fully developed turbulence, 
the energy should be transferred downwards from large scale to small
scales following a
self-similar and homogeneous process entirely dependent
on the energy transfer rate, $\epsilon$, and the scale, $l$. 
By assuming local homogeneity and isotropy, the K41 theory allows us to  
predict the
scaling properties of the structure functions
$S_p(l) \equiv \langle(\delta v(l))^p\rangle$, where $\delta v (l) \sim  
|v(x+l)-v(x)|$.
It turns out that 
$S_p(l) \sim 
\langle \epsilon \rangle ^{p/3} l^{p/3}$. 
The K41 theory has been questioned by several authors because  of 
strong scale dependent fluctuations of the energy dissipation  
(intermittency).
Because of intermittency, the scaling properties of the velocity  
structure functions
acquire anomalous scaling, i.e., $S_p(l)\sim l^{\zeta(p)}$, where the  
scaling exponents
$\zeta (p)$ are non linear functions of $p$ \cite{MS,BCTBS}. 

The universal character of
energy transfer statistics has been also questioned. Different
 $\zeta(p)$ exponents have been measured in anisotropic
flows (shear flows \cite{roberto_shear} and boundary layers 
\cite{sreeni_bl})
and strong dependencies on  active quantities as the temperature
in Rayleigh-Benard \cite{roberto_rayleigh} or the magnetic field in MHD
\cite{MHD} have been quoted. \\ 
The question of
finding unifying and universal aspects common to all turbulent
flows naturally arises.\\
Recently,
cascade descriptions based on random multiplicative processes and
infinitely divisible distributions for random multipliers  have been 
applied in order to clarify
the energy transfer physics.
In particular  it has been recently shown 
(\cite{sl}-\cite{sw}) that a log-Poisson distribution is able to provide  
an 
extremely good fit to experimental data.
In \cite{sw} authors proposed that
the random multipliers $W_{L_2,L_1}$ connecting velocity fluctuations at
different scales $L_2$ and $L_1$
\begin{equation}
\delta v(L_2) =  W_{L_2,L_1} \delta v(L_1)
\label{multiplier}
\end{equation}
should follow a log-poisson statistics.\\  
Log-poissonity should naturally arise in turbulent flows as the limit
of a Bernoulli fragmentation process between 
infinitely close scales. In \cite{sw} authors argued that 
the energy   transfer between
two adjacent scales $l_1$ and $l_2$ with $\log(l_1)-\log(l_2) << 1$
can be described in terms of only two bricks. The first one corresponds
to  the most 
 singular structure  characterized by a local scaling exponents
$h_0$:  $ \delta v(l_2) = (l_2/l_1)^{h_0} \delta v(l_1)$.
The second brick 
 is  a {\it defect}-like  energy transfer which modulates the most singular
events by a factor $\beta < 1$; in this second case we would have
a typical scaling: $ \delta v(l_2) = \beta (l_2/l_1)^{h_0} \delta v(l_1).$
Let us also assume that, in the limit $\log(l_1/l_2) \rightarrow 0$,
 {\it defects}  along the cascade happen with 
a probability that goes to zero proportionally to the logarithm of the
scale separation $ \sim d_0 \log(l_1/l_2)$,
where $d_0$ is the parameter  which controls how probable a  defect is.
It is now  simple to show, following \cite{sw}, that the 
finite-scale-separation 
transfer must have a log-poisson statistics: 
$<(W_{L_2,L_1})^p> = (L_2/L_1)^{h_0p -d_0(1-\beta^p)}$,
which corresponds to 
 the  She-Leveque proposal for intermittent exponents \cite{sl}:
\begin{equation}
\zeta(p) = h_0 p -d_0 (1-\beta^p).
\label{sheleveque}
\end{equation}
The  multifractal interpretation
of (\ref{sheleveque}) is that $h_0$ is the most singular scaling
 exponent and  $d_0$
corresponds to 
 the codimension of the fractal set where the most singular scaling
is observed.
One may argue that  the structure and the statistics
of the most singular event could be strongly non-universal.\\
As a consequence, a possible scenario
can be proposed where  $h_0 $ and $d_0$ 
 could be system-dependent, while $\beta$ maybe
constant in a wider universality class \cite{sw}. \\
Some evidences of this universal character of $\beta$ 
have already been reported in \cite{gess1} where  it
has been shown that the differences of intermittent exponents
measured in Rayleigh-Benard, MHD, shear flows
and in boundary layers can all be re-adsorbed  by properly changing $h_0$
and $d_0$
at constant $\beta$. In \cite{gess1} it has been shown
that also viscous effects can be included in a suitable 
dependency on the  scale of $h_0$ and $d_0$, showing for the first time
that non-trivial intermittent corrections, due to the $\beta$ dependent
part of $\zeta(p)$ curve, can be detected at viscous scales.

In the   following we will show how log-poisson description
of intermittency and its underlying interpretation
can be  applied for describing some important aspects
of energy transfer in a class of dynamical models of turbulence 
\cite{leo_physicstoday,bbkt,BK}.
In particular, we  will be able to  explain the
continuous transition towards the K41 non-intermittent  statistics
in terms of the statistical properties
of the most singular structure of the model.  This trend 
towards  K41 statistics is observed in a class of shell models
at varying (diminishing) the characteristic
 shell separation, i.e. in the so-called
continuous limit of infinitely close shells.\\
Universal aspects of the shell statistics are recovered by 
 measuring the $\beta$-dependent part of the probability 
distribution. As proposed by She and Waymire, we find that 
the $\beta$-dependent part of the intermittent statistics
is remarkably constant for all values of the shell
separation $\lambda$ explored.

Shell models have demonstrated to be very useful for the understanding
of many properties connected to the non-linear turbulent energy
transfer (\cite{BK}-\cite{BKLMW}).
The most popular shell model, the Gledzer-Ohkitani-Yamada 
(GOY) model (\cite{BK}-\cite{BKLMW}), has been shown to predict
scaling properties for $\zeta(p)$ similar to what is found experimentally
(for a suitable  choice of the free
parameters).  

The GOY model can be seen as a severe truncation of the 
Navier-Stokes equations: 
it retains only one complex mode $u_n$ as a representative 
of all Fourier modes in 
the shell of wave numbers $k$ between $k_n=k_0\lambda^n$ 
and $k_{n+1}$, $\lambda$
being an arbitrary scale parameter ($\lambda>1$), usually 
taken equal to $2$. 

 In two recent works \cite{BK,bbkt} the GOY
model has been generalized in terms of shell variables, $u_n^+$ and
$ u_n^-$,
transporting positive and negative helicity, respectively.  These models have 
at least one inviscid invariant non-positive defined which is
very similar to the 3d Navier-Stokes helicity.  In the following we will
focus on the intermittent properties of one of such models at varying
the separation between shells, $\lambda$. The time evolution
for positive-helicity shells is \cite{bbkt}:
\begin{eqnarray}
\frac{d}{dt} u^+_n&=&  
i k_n ( u^{-}_{n+2} u^{+}_{n+1}+b 
u^{-}_{n+1} u^{+}_{n-1}
+c u^{-}_{n-1} u^{-}_{n-2})^*\nonumber \\
&-& \nu k_n^2 u^+_n  +\delta_{n,n_0} f^+,
\label{eq:shells}
\end{eqnarray}
and the same, but with all helical signs reversed, holds for $u_n^-$.

The coefficients $b,c$ are determined imposing  inviscid 
conservation of energy, $E= \sum_n(\vert u^+_n \vert^2 +
 \vert u^-_n \vert^2))$,  and helicity, $H=\sum_n k_n
(\vert u^+_n \vert^2 - \vert u^-_n \vert^2)$.
 In particular we have: $b=-(1+\lambda^2)/(\lambda^3+\lambda^2)$
 and $c=(1-\lambda)/(\lambda63+\lambda64)$.\\
Structure functions for these models are naturally defined as
$ S_p(n) = <~(\sqrt{|u_n^+|^2 + |u_n^-|^2})^p> \sim k_n^{-\zeta(p)}.$
Many investigation have been done on the intermittent properties
of this model \cite{bbkt} and of the original GOY model 
\cite{JPV,BLLP,BKLMW} at varying the coefficients
such as to have different conserved quantities, in order  to investigate 
the importance of inviscid invariants in determining the energy transfer
properties. Some evidences supporting non-trivial effects introduced
by non-positive invariants have been quoted.  \\
In this letter
we discuss the dependency on the other important free parameter
entering shell-modelization: the separation between neighboring shells
$\lambda$.  By decreasing $\lambda$ one describes
 turbulent energy transfer in terms of interactions which become more
and more local in Fourier space. In the limit of $\lambda \rightarrow 1$ 
one recovers 
a $1$-dimensional partial differential equation \cite{parisi,bbkt}. 
 Locality of interactions
has always been a long-debated issues of the K41 picture.
In fig. 1 we show the $6$th order structure function of model (\ref{eq:shells}) 
for different inter-shell separations: $\lambda=2$ and $\lambda=1.05$.
In both cases
  the total number of shells is chosen  such  that the 
physical length of the inertial range stays constant. Clearly, there is a 
net trend toward a less intermittent state by decreasing the shell-ratio.
Nevertheless,
the exceptionally good scaling properties allow us to estimate,
using the Extended Self Similarity method \cite{BCTBS},
small deviations from K41 scaling also for the less intermittent value
$\lambda=1.05$.

In view of the discussion about log-poisson statistics this analysis
could not be sufficient. Non-universal aspects can be masked by
trivial  properties of the most singular structure. 
We have therefore tried to highlight possible
universal aspects of the chaotic energy transfer by focalizing our
attention on the non-linear part of intermittent corrections and in particular
on the parameter $\beta$ which can be extracted by assuming log-poisson
intermittency. The best way of extracting  $\beta$ from numerical data
is to look if structure functions verify the log-poisson 
hierarchy \cite{sl}:
\begin{equation}
\frac{S_{p+1}(n)}{S_p(n)} = A_n [\frac{S_p(n)}{S_{p-1}(n)}] ^{\beta^\prime},
\label{eq:hierarchy}
\end{equation}
where $\beta^\prime=\beta^{1/3}$ and $A_n$ has a sca\-ling de\-pen\-dency 
on $k_n$ 
but it is $p$-in\-depen\-dent. If structure functions  follow
a log-poisson statistics  one can extract $\beta^\prime$ from a linear
fit of a log-log plot of (\ref{eq:hierarchy}), at varying $p$ and fixed 
$n$.\\
In fig. 2 we show relation (\ref{eq:hierarchy}) for the two 
inter-shell separations  $\lambda=2,1.05$ and for $p=1,...,7$.
 As it is possible to see, the straight line behaviour
in the log-log plot is nicely verified (supporting the log-poisson assumption)
and the slope of the two lines is the same, indicating that although
the full intermittent corrections are different, the part directly
affected by the  $\beta$ parameter is the same.  In table 1 we collect
all our estimates of $\beta^\prime$ as a function of $\lambda$. From these data
one can safely conclude that   $\beta$ stays constant by approaching 
the continuous limit $\lambda \rightarrow 1$. The remarkable 
change in the statistics reflected by changes in the 
 $\zeta(p)$ exponents of  fig. 1 may be only due to changes
of the most singular energy-burst statistics,  i.e.,  
 to changes of the parameters
$h_0 $ and $d_0$ in the log-poisson jargon.
Let us notice that already other authors 
 have focused the attention on burst-like solutions in GOY models as possible 
explanation of its intermittent properties
\cite{parisi,dombre}. Moreover, it is very simple to interpret the 
measured tendency
toward K41 statistics as the results
of a smoothing in the energy transfer due to the increasing
local character of dynamical interactions, in the limit 
$\lambda \rightarrow 1$. In 
other words, we interpret the corrections to K41 for $\lambda >> 1$ as
the consequence of burst-structures travelling along
 the inertial range. Very singular bursts
(with $h_0 $ less than the Kolmogorov value $1/3$)
appear only when there is a relevant mismatch between eddy-turnover times
of neighboring shells, i.e.
 only when shell-ratios are much larger than one. Otherwise,
energy tends to flow smoothly toward small scales following K41 statistics.\\
These bursts are
non-universal in the sense that their statistics and their degree of
singularity is a function of the inter-shell ratio, $\lambda$.

On the other hand,  the mechanism underlying 
 burst transfer seems only determined by the non-linear structure
of shell-model dynamics and it seems to fix a universal
value of $\beta$ in the log-poisson description.

To conclude, we have analysed and discussed the intermittent properties of
a class of shell models for turbulent energy transfer at varying 
the inter-shell ratio parameter. 
We have shown that there exists a suitable limit,
$\lambda \rightarrow 1$, for which intermittent
fluctuations disappear and the scaling properties of
the system become close to the K41 theory.
By employing a log-poisson description of
intermittent corrections, we have disentangled 
the limit $\lambda \rightarrow 1$ in two main components:
from one hand the most singular structure seems to tend
continuously toward a K41 scaling ($h_0=1/3$); from
the other hand the probability distribution of intermittent
fluctuations is still log-Poisson and described by the same
{\it defect} parameter $\beta$. One possible way 
of describing this result, in terms
of the scaling exponents $\zeta(p)$, consists in saying that the 
quantity

\begin{equation}
\rho_{pq} =\frac{\zeta(p)-p/3 \,\zeta(3)}{\zeta(q)-q/3 \,\zeta(3)} 
\end{equation}
 
is constant for $\lambda \rightarrow 1$. Similar universal properties
have already been reported in shell models at varying the dissipation
mechanism, i.e. using hyperviscosities \cite{sl_hypervis}.\\
Our results are qualitatively similar to what observed in real turbulent
flows  going from the inertial subrange to the viscous subrange. Indeed
in this case anomalous scaling disappears (i.e. $\zeta(p)/\zeta(3) 
\rightarrow p/3$) while $\rho_{pq}$
stays constant, as recently observed in \cite{gess1}.

\centerline{FIGURE CAPTIONS}

\begin{itemize}

\item FIGURE 1: Log-log plot of the 6th order structure function vs
 $k$, for the two cases $\lambda=2$
 (circles) and $\lambda=1.05$ (diamonds).
 A linear fit in the inertial range, using Extended Self Similarity,
 gives: $\zeta(6)/\zeta(3)=1.983\pm0.005$ for $\lambda=1.05$ and 
 $\zeta(6)/\zeta(3)=1.76\pm0.01$ for $\lambda=2$.
\item FIGURE 2: Log-poisson hierarchy (eq.~\ref{eq:hierarchy}) in a 
 log-log plot for $\lambda=2$ (circles) and $\lambda=1.05$ (diamonds).
 For each $\lambda$, we took $p=1,...,7$ for three different scales 
 $n$ in the inertial range and we shifted the sets
 along y-axis in order to perform 
 a single linear fit (solid lines).
\end{itemize}

\centerline{TABLE CAPTIONS}

\begin{itemize}

\item TABLE 1:  $\beta^\prime$ at varying $\lambda$.
 Each value is the average  
 of slopes evaluated as in fig. 2 
 for all scales $n$ in the inertial range. 
 The theoretical prediction \cite{sl} is
 $\beta^\prime=(2/3)^{1/3}=0.87$.
\end{itemize}


%
\begin{table}
\begin{center}
\begin{tabular}{|c|c|}
\hline
$\lambda$ & $\beta^\prime$\\
\hline
$1.05$ & $0.86 \pm 0.01$\\
$1.2$ & $0.88 \pm 0.01$\\
$1.5$ & $0.88 \pm 0.01$\\
$2$ & $0.86 \pm 0.01$\\
\hline
\end{tabular}
\end{center}
\label{tab}
\end{table}
%


\begin{thebibliography}{99}

  \bibitem{MS} C. Meneveau and K. R. Sreenivasan, J. Fluid Mech. 
  {\bf 224}, 429 (1991).

  \bibitem{BCTBS} R. Benzi, S. Ciliberto, R. Tripiccione, C. Baudet, 
  C. Massaioli and S. Succi, Phys. Rev. E {\bf 48}, R29 (1993);
   R. Benzi, S. Ciliberto, C. Baudet and G. R. Chavarria, 
   Physica D {\bf 80}, 385 (1994).

  \bibitem{roberto_shear} R. Benzi, S. Ciliberto, C. Baudet, G. Ruiz 
   Chavarria, R. Tripiccione, Europhys. Lett. {\bf 24}, 275 (1993).

  \bibitem{sreeni_bl} G. Stolovitzky, K. R. Sreenivasan, Phys. Rev. E 
  {\bf 48}, 32 (1993).

  \bibitem{roberto_rayleigh} R. Benzi, R. Tripiccione, F. Massaioli, S. Succi
   and S. Ciliberto, Europhys. Lett. {\bf 25}, 331 (1994).

  \bibitem{MHD} R. Grauer, Phys. Lett. A {\bf 195}, 335 (1994).

  \bibitem{sl} Z. S. She and E. Leveque, Phys. Rev. Lett. {\bf 72}, 
   336 (1994).

  \bibitem{d} B. Dubrulle, Phys. Rev. Lett. {\bf 73}, 959 (1994).

  \bibitem{novikov} E.A. Novikov, Phys. Rev. E {\bf 50}, R3303 (1994).

 \bibitem{sw} Z.S. She and E.C. Waymire, Phys. Rev. Lett. {\bf 74}, 262 (1995).
  
  \bibitem{gess1} R. Benzi, L. Biferale, S. Ciliberto,
  M. V. Struglia and R. Tripiccione, Phys. Rev. E {\bf 53}, 3025 (1996).  

  \bibitem{leo_physicstoday} L. Kadanoff, 
   Physics Today {\bf 48}, 11 (1995).

  \bibitem{BK} L. Biferale and R. Kerr,  Phys. Rev. E  {\bf 52}, 6113 (1995).  

  \bibitem{bbkt} R. Benzi, L. Biferale, R. Kerr and E. Trovatore,
   Phys. Rev. E  {\bf 53}, 3541 (1996). 

  \bibitem{G} E.B. Gledzer, Sov. Phys. Dokl. {\bf 18}, 216 (1973).
  
  \bibitem{YO} M. Yamada and K. Ohkitani, Prog. Theor. Phys. {\bf 81},
  329 (1989); J. Phys. Soc. Jpn. {\bf 56}, 4210
  (1987); Phys. Rev. Lett {\bf 60}, 983 (1988). 

  \bibitem{JPV} M.H. Jensen, G. Paladin and A. Vulpiani, 
  Phys. Rev. A {\bf 43}, 798 (1991).

  \bibitem{BLLP} L. Biferale, A. Lambert, R. Lima and G. Paladin, Physica D
  {\bf 80}, 105 (1995).

  \bibitem{PBCFV} D. Pisarenko, L. Biferale, D. Courvoisier, U. Frisch and
  M. Vergassola, Phys. Fluids {\bf A65}, 2533 (1993).

  \bibitem{BKLMW} L. Kadanoff, D. Lohse, J. Wang and R. Benzi,
  Phys. Fluids {\bf 7}, 617 (1995).
  
  \bibitem{parisi} G. Parisi, unpublished (1990). 

  \bibitem{dombre} T. Dombre and J. L. Gilson,
  {\it Intermittency, chaos and singular fluctuations in the mixed 
  Obukhov-Novikov shell model of turbulence}, preprint (1996). 

  \bibitem{sl_hypervis} Z. S. She and E. Leveque, Phys. Rev. Lett. 
  {\bf 75}, 2690 (1995).


  \end{thebibliography}
\end{document}